\documentclass[prl,reprint,superscriptaddress,aps,floatfix]{revtex4-2}
\usepackage{amsmath}
\usepackage{amsfonts}
\usepackage{graphicx}
\usepackage{physics}
\usepackage[usenames]{color}
\usepackage{braket}
\usepackage{nicematrix, arydshln}
\usepackage{physics}
\usepackage{tabularray}
\usepackage{array}
\usepackage{url}
\usepackage[colorlinks=true, allcolors=blue]{hyperref}
\usepackage{cleveref}
\usepackage[dvipsnames]{xcolor}
\usepackage{subcaption}
\usepackage{ragged2e}
\usepackage{tikz}

\begin{document}
\title{Tunably realizing flat-bands and exceptional points in kinetically frustrated systems: An example on the non-Hermitian Creutz ladder}
\author{Debashish Dutta}
\author{Sayan Choudhury}
\email{sayanchoudhury@hri.res.in}
\affiliation{Harish-Chandra Research Institute, a CI of Homi Bhabha National Institute, Chhatnag Road, Jhunsi, Allahabad 211019}

%%%%%%%%%%%%%%%%%%%%%%%%%%%%%%%%%%%%%%%%%%%%%%%%%%%%%%%%%%%%%%%%%%%%
%New section
%%%%%%%%%%%%%%%%%%%%%%%%%%%%%%%%%%%%%%%%%%%%%%%%%%%%%%%%%%%%%%%%%%%%

\begin{abstract}
We study a non-Hermitian extension of the Creutz ladder with generic non-reciprocal hopping model. By mapping the ladder onto two decoupled non-Hermitian SSH chains, we uncover a rich parameter-space structure under different boundary conditions. Under periodic boundary conditions (PBC) the spectrum admits a fine-tuned line in parameter space with entirely real eigenvalues, while deviations from this line induce a real–complex spectral transition without crossing exceptional points. In contrast, an exact analytical diagonalization under open boundary conditions (OBC) reveals extended regions in parameter space with purely real or purely imaginary spectra, separated from complex spectral domains by exceptional lines.
The intersections of these exceptional lines define triple-junction points where distinct spectral regimes meet, giving rise to a structured phase diagram absent under periodic boundaries. We show that flat bands in this system can occur both as Hermitian diabolical points and as non-Hermitian exceptional points, known as exceptional flat bands, where the dynamics is more stringent than its Hermitian counterpart, leading to distinct spectral and dynamical signatures.
\end{abstract}

\maketitle

%%%%%%%%%%%%%%%%%%%%%%%%%%%%%%%%%%%%%%%%%%%%%%%%%%%%%%%%%%%%%%%%%%%%
%New section
%%%%%%%%%%%%%%%%%%%%%%%%%%%%%%%%%%%%%%%%%%%%%%%%%%%%%%%%%%%%%%%%%%%%

\section{Introduction}
Non-Hermitian Hamiltonians have become a central framework for describing open, dissipative, and non-reciprocal systems \cite{open1,open2,open3,open4}, giving rise to spectral and dynamical features absent in Hermitian quantum mechanics. 
In contrast to Hermitian operators, which guarantee real spectra and orthogonal eigenstates, non-Hermitian models can exhibit complex eigenvalue spectra and real–complex spectral transitions \cite{bender}. 
The quintessence of non-Hermitian extension of quantum mechanics lies in generalizing the existing framework to reveal deeper physics and more intricate mathematical structures. This includes a nuanced topological classification of Hamiltonians \cite{NHtopo5,NHtopo1,NHtopo2,NHtopo4}, exceptional point degeneracies \cite{EP1,EP2,EP3,EP4,EP5,banerjee2023tropical}, complex band gaps \cite{NH-review1,NHtopo2} and non-Bloch band theory \cite{NBBT1,NBBT2,NBBT3,NBBT4,NBBT5,NBBT6,NBBT7,NBBT8}.
Non-Hermitian systems exhibit strong boundary sensitivity, exemplified by the non-Hermitian skin effect (NHSE) \cite{NHSE1,NHSE2,NHSE3,NHSE4,NHSE5,NHSE6,NHSE7,NHSE8}, where a macroscopic number of eigenstates localize at system edges under open boundary conditions, resulting in the breakdown of conventional bulk–boundary correspondence \cite{BBC1,BBC2,BBC3}.
Collectively, these features make non-Hermitian systems promising platforms for further investigation and a fertile ground for quantum engineering.

On the other hand, flat-band (FB) systems are characterized by their quenched kinetic energy, leading to macroscopic spectral degeneracies \cite{fb1,fb2,fb3,fb4,fb5,fb6}. 
Geometric frustration\cite{gf1,gf2} implicit in their lattice architecture enforces destructive interference of hopping amplitudes from different channels, causing compact localization (CL) \cite{cl1,cl2,cl3,cl4,cl5,cl6} of wave packets.
As a consequence, even weak perturbations or interactions can become relevant, making flat-band systems suitable platforms for strongly correlated phases such as superconductivity \cite{fb-sc1,fb-sc2,fb-sc3}, ferromagnetism \cite{fb-mag1,fb-mag2,fb-mag3}, quantum Hall states \cite{fb-hall1,fb-hall2,fb-hall3}, and Wigner crystallization \cite{fb-wc1,fb-wc2}.

The Creutz ladder constitutes a paradigmatic example of such a flat-band system, featuring two exactly flat bands at fine-tuned hopping parameters and exhibiting robust compact localization protected by lattice geometry \cite{creutz1,creutz2}.
Owing to its simple structure and exact solvability in the Hermitian limit, the Creutz ladder has long served as a canonical platform for studying flat-band physics, disorder-free localization, and boundary effects.
Introducing non-Hermiticity into this model raises a number of fundamental questions: how flat-band degeneracies reorganize in the presence of non-reciprocal hopping, whether flat bands persist as Hermitian diabolical points or acquire exceptional-point character, and how boundary sensitivity manifest in a frustrated geometry.
These features make the non-Hermitian Creutz ladder an ideal testbed for analytically probing the interplay between flat-band physics, exceptional degeneracies, and boundary-condition-dependent spectral behavior.
Non-Hermitian extensions of flat-band systems and ladder models have been explored in a variety of contexts, including gain–loss engineering, asymmetric hopping, and photonic realizations, revealing exceptional degeneracies and boundary-sensitive spectral features \cite{nhfb1,nhfb2,nhfb3,nhfb4,qi2018defect,banerjee2024non,martinez2024topological}.

In view of the growing interest in non-Hermitian flat-band systems and ladder geometries, it is natural to seek analytically tractable models that allow a systematic comparison of spectral and boundary-condition-dependent properties. In particular, an explicit characterization of flat-band degeneracies and their evolution under non-reciprocal hopping, as well as a clear understanding of how periodic and open boundary spectra relate in such systems, remain valuable from both conceptual and practical perspectives. In this work, we present an analytical study of a non-Hermitian extension of the Creutz ladder with generic non-reciprocal hopping. By exactly diagonalizing the model under both periodic and open boundary conditions and exploiting its mapping to decoupled non-Hermitian Su–Schrieffer–Heeger (SSH) chains, we elucidate the structure of flat bands, boundary-sensitive spectra, and their organization in parameter space. The analytical tractability of the model further enables a direct connection between spectral properties and wave-packet dynamics, providing a unified perspective on non-Hermitian flat-band physics in ladder systems.

The remainder of the paper is organized as follows. In Sec. II, we introduce the non-Hermitian Creutz ladder with non-reciprocal hopping and analyze the spectrum under periodic boundary conditions and identify the conditions for flat-band formation. Sec. III presents an exact analytical diagonalization of the model under open boundary conditions and examines the resulting spectral structure and exceptional degeneracies. In Sec. IV we investigate boundary localization of eigen states and Sec. V is dedicated towards the study of wave-packet dynamics under different parameter conditions in the non-Hermitian regime. We conclude in Sec. VI with a summary and outlook.
\label{sec:Introduction}

%%%%%%%%%%%%%%%%%%%%%%%%%%%%%%%%%%%%%%%%%%%%%%%%%%%%%%%%%%%%%%%%%%%%
%New section
%%%%%%%%%%%%%%%%%%%%%%%%%%%%%%%%%%%%%%%%%%%%%%%%%%%%%%%%%%%%%%%%%%%%

\section{Model and PBC spectrum}
\begin{figure}
    \centering
    \includegraphics[width=\linewidth]{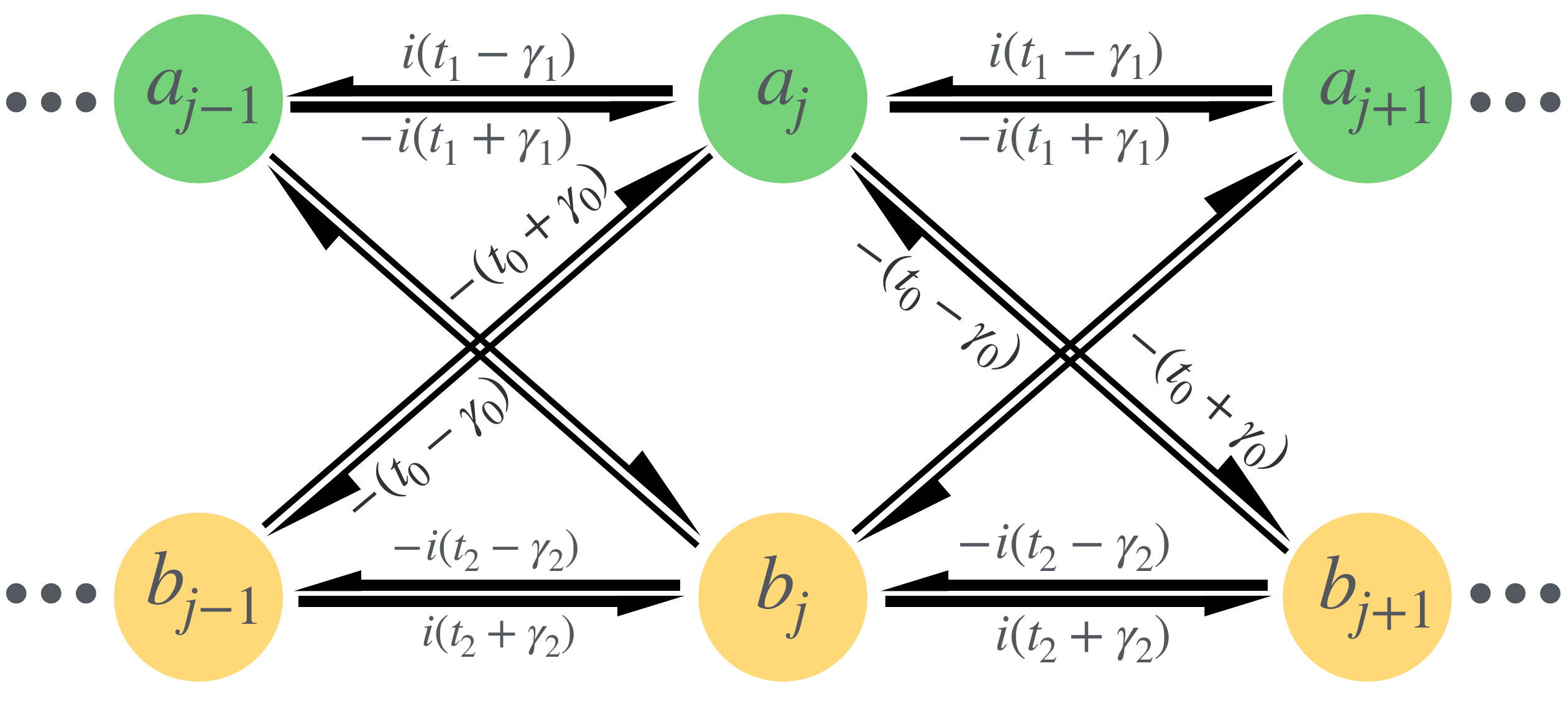}
    \caption{
    \small
    \justifying
    \textbf{Non-Hermitian Creutz ladder with non-reciprocal hopping.}
    Schematic of the ladder geometry consisting of two cross-linked chains. Sublattice points of the upper (lower) chain are represented with green (yellow) circles and labeled with $a_j \, (b_j)$, where $j$ is the cell index. The model is rendered non-Hermitian by pairing reciprocal hoppings $(\{t\})$ with non-reciprocal terms $(\{\gamma\})$. The arrows represent the hoppings from one lattice point to another.}
    \label{fig:lattice diagram}
\end{figure}
In the Hermitian Creutz ladder, two lattice chains are cross-linked by the hopping amplitude \(t_0\), while \(t_1\) and \(t_2\) denote the inter-chain hopping strengths along the upper and lower legs, respectively. At the fine-tuned parameter values \(t_1 = t_2 = \pm t_0\), the lattice geometry enforces incompatible hopping amplitudes that suppress kinetic dispersion and generate a completely degenerate double flat-band spectrum at \(E = \pm 2t_0\). The resulting eigenstates form compactons whose dynamics remains strictly confined to a pair of adjacent plaquettes due to destructive interference between alternative hopping paths, leading to compact localization (CL).

We consider a non-Hermitian generalization of the Creutz ladder obtained by augmenting each reciprocal hopping process with a corresponding non-reciprocal component. The resulting lattice geometry is illustrated in \cref{fig:lattice diagram}, and the system is governed by the Hamiltonian
\begin{figure}[t]
    \centering
    \includegraphics[width=\linewidth]{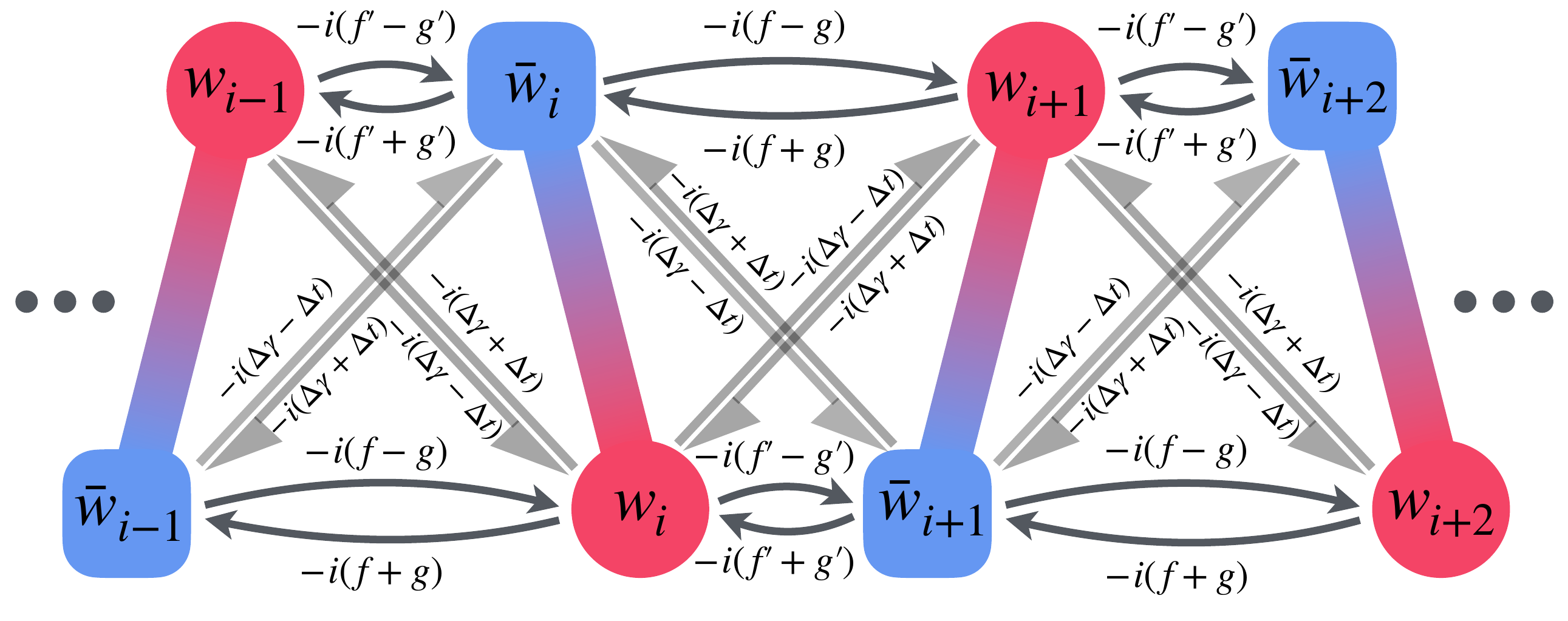}
    \caption{
    \small
    \justifying
    \textbf{Mapping of the non-Hermitian Creutz ladder to cross-coupled NH-SSH chains.} After a local basis transformation, the non-Hermitian Creutz ladder can be represented as two cross-linked non-Hermitian Su–Schrieffer–Heeger (NH-SSH) chains composed of $w$ and $\bar{w}$ sites, shown in red circles and blue squares respectively, with a bridge pairing them highlighting that they belonged to the same cell in the original ladder. For balanced intra-chain hoppings $(\Delta t= \Delta \gamma=0)$, the two chains decouple exactly.
    }
    \label{fig:Double SSH lattice}
\end{figure}

\begin{equation}
\begin{aligned}
    \mathcal{H} = \sum_{j} 
    & -i(t_1+\gamma_1)a^\dagger_{j+1}\, a_j + i(t_2+\gamma_2) b^\dagger_{j+1}\,b_j \\
    & +i(t_1-\gamma_1)a^\dagger_j \, a_{j+1} -i(t_2-\gamma_2) b^\dagger_j\,b_{j+1} 
    \\
    & 
    -(t_0+\gamma_0) (a^\dagger_{j+1}\,b_j + b^\dagger_{j+1}\,a_j)
    \\
    &
    -(t_0-\gamma_0) (a^\dagger_j\,b_{j+1} + b^\dagger_j\,a_{j+1})
\end{aligned}
\end{equation}

The $a^{(\dagger)}_j,b^{(\dagger)}_j$ are the fermionic annihilation (creation) operators acting on the j-th cell of the upper and lower lattice chains respectively. The reciprocal hopping terms $t_1,t_2,t_0$ are now paired with the non-reciprocal $\gamma_1,\gamma_2,\gamma_0$ parameters.

\textit{PBC spectrum.} The PBC spectrum can easily be studied by going to the reciprocal space where it block diagonalizes as $\mathcal{H} = \Psi^\dagger_k h(k) \Psi_k$, where $\Psi_k = (a_k,b_k)^\text{T}$ and

\begin{equation}
    h(k) = 2
    \begin{pmatrix}
        t_1\sin k -i\gamma_1 \cos k  &  -t_0 \cos k -i\gamma_0 \sin k \\
        -t_0 \cos k -i\gamma_0 \sin k  &  -t_2\sin k + i\gamma_2 \cos k
    \end{pmatrix}
    \label{eq:hk}
\end{equation}

with the dispersion relation being
\begin{equation}
\begin{aligned}
    E^{\text{PBC}}_\pm(k) = 
    &
    2 (\Delta t\sin k -i\Delta \gamma \cos k) 
    \\ \pm  
    &
    2\sqrt{
    \begin{aligned}
    &(t_0^2-\bar{\gamma}^2)\cos^2k +(\bar{t}^2-\gamma_0^2)\sin^2 k 
    \\
    &+i(t_0\gamma_0 - \bar{t}\bar{\gamma})\sin 2k
    \end{aligned}
    }
\end{aligned}
\end{equation}

 Here we have redefined the parameters as $\bar{t}=(t_1+t_2)/2$, $\bar{\gamma}=(\gamma_1+\gamma_2)/2$, $\Delta t= (t_1-t_2)/2$ and $\Delta \gamma = (\gamma_1 - \gamma_2)/2$.

In this work, we focus on the regime of balanced intra-chain hoppings, corresponding to $\Delta t = \Delta \gamma = 0$. We fix the energy scale by setting $\bar{t} = 1$, while treating $t_0$ and $\bar{\gamma}$ as continuously tunable parameters and varying $\gamma_0$ discretely.

The PBC spectrum of this model has the loop structures in the complex energy plane as generic to many non-Hermitian systems.
But the system hosts a fine tuned line $t_0 \, \gamma_0 =  \bar{t} \, \bar{\gamma}$ in the $(\bar{t},\gamma_0)$ parameter space where the eigen spectrum is entirely real when $\bar{t} > \gamma_0$, see \cref{fig:spectral domain numerical}(a,b). 
Along this line the spectrum takes the form $E^{\text{PBC}}_\pm(k) = \pm 2\sqrt{\bar{t}^2-\gamma_0^2} \sqrt{ \eta^2 \cos ^2k + \sin ^2 k }$, where $\eta$ parametrizes the line from the origin.
This dispersion relation is analogous to a Hermitian Creutz ladder where $\sqrt{\bar{t}^2-\gamma_0^2}$ acts as intra-chain hopping $\bar t$ and $\eta$ corresponds to $t_0/\bar t$.
Away from this line, the spectrum becomes complex, resulting in a real-to-complex spectral transition mediated by Hermitian degeneracies or diabolical points (DPs), rather than by exceptional points (EPs).
At $\eta=\pm 1$, the two bands collapse into dispersionless double flat bands (DFBs) supporting compactly localized eigenstates. 
Since the Hamiltonian remains diagonalizable, these flat bands correspond to diabolical-point degeneracies in a non-Hermitian setting.
By contrast, when \(\bar{t}=\pm\gamma_0\), the two flat bands merge at \(E=0\) and the Hamiltonian becomes defective, forming Jordan blocks of order two for all values of \(\eta\).
This complete degeneracy along an exceptional line of second order (EL2) gives rise to an exceptional flat band (EFB) \cite{efb1}. 
Flat bands formed at exceptional points are therefore associated with more relaxed tuning constraints, as illustrated here by an entire line in parameter space supporting a flat band.
For \(|\gamma_0|>|\bar{t}|\), the flat band persists but acquires a finite lifetime due to the emergence of imaginary energy components.
\label{sec:Model and Hamiltonian}

%%%%%%%%%%%%%%%%%%%%%%%%%%%%%%%%%%%%%%%%%%%%%%%%%%%%%%%%%%%%%%%%%%%%
%New section
%%%%%%%%%%%%%%%%%%%%%%%%%%%%%%%%%%%%%%%%%%%%%%%%%%%%%%%%%%%%%%%%%%%%

\section{OBC Spectrum and Exceptional lines}
\begin{figure*}
    \centering
    \includegraphics[width=\linewidth]{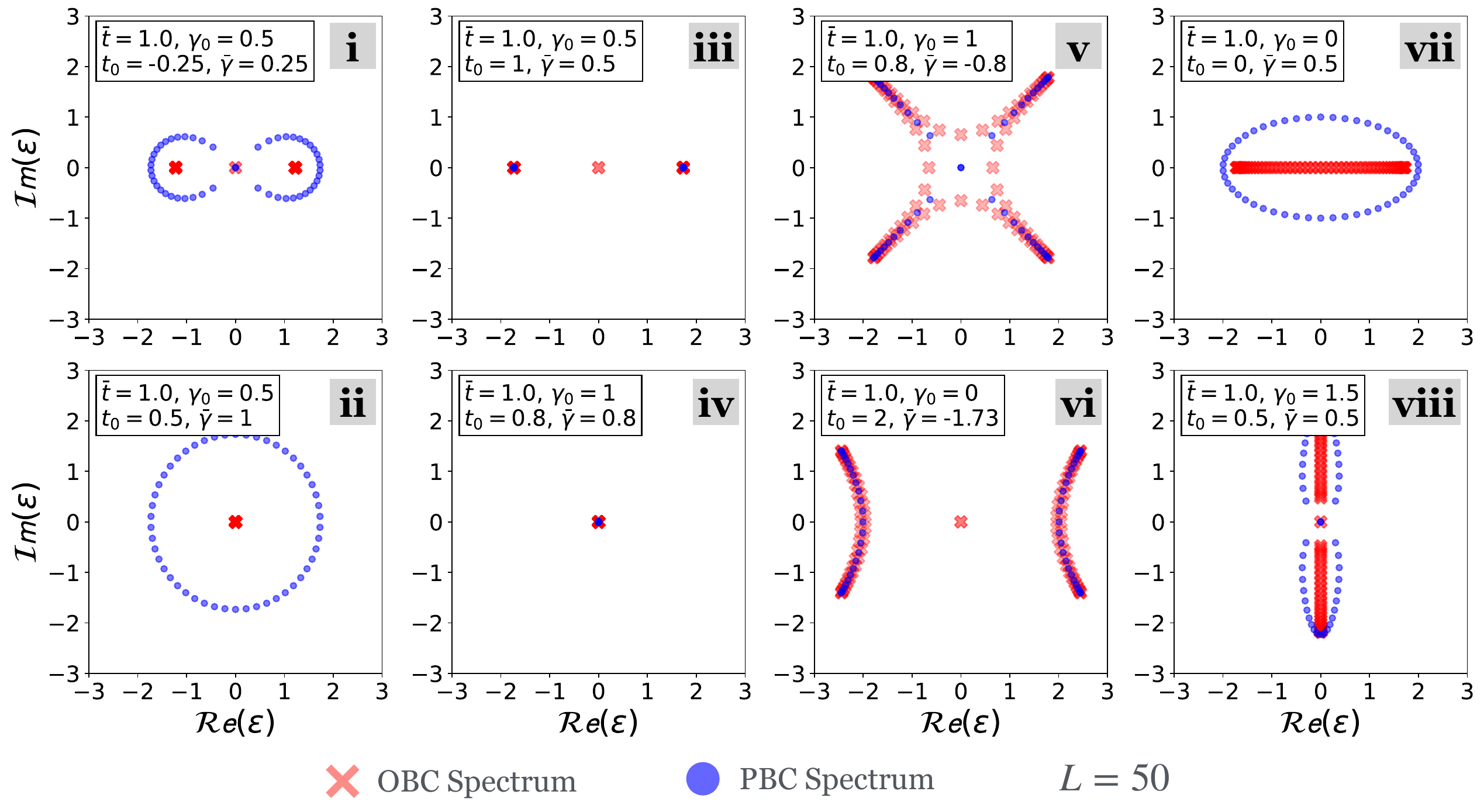}
    \caption{
    \small
    \justifying
    \textbf{Comparison of PBC and OBC spectra of the non-Hermitian Creutz ladder.}
    Complex-energy spectra under PBC (blue dots) and OBC (red crosses) for representative parameter values. Panels (i–ii) illustrate dispersive PBC and degenerate OBC spectra. Panels (iii-iv) shows flat band PBC and degenerate OBC spectra, while panels (v–vi) show cases where the PBC spectrum encloses no area in the complex plane, leading to the disappearance of the non-Hermitian skin effect and restoration of bulk–boundary correspondence. Panels (vii–viii) demonstrate regimes where the OBC spectrum is entirely real or imaginary despite a complex PBC spectrum, highlighting the strong boundary sensitivity of non-Hermitian systems.
    }
    \label{fig:spectrum plots}
\end{figure*}

Creutz model simplifies when we make a local transformation from $(a,b)$ basis to $(w,\bar{w})$ basis \cite{creutz2}, defined by

\begin{align}
    \nonumber
    w_i &= (a_i + ib_i )/\sqrt{2}
    \\
    \bar{w}_i &= (a_i - ib_i )/\sqrt{2}
\end{align}

With a suitable rearrangement of these bases it can be shown that this model is equivalent to two non-Hermitian SSH (NH-SSH) chains with non-reciprocal nearest neighbour (NN) hoppings, cross linked together by parameters that depend only on $\Delta t$ and $\Delta \gamma$, as shown in \cref{fig:Double SSH lattice}. For the case we are considering, where $\Delta t,\Delta\gamma=0$, the two chains decouple and the Hamiltonian splits into two parts $H_1$ and $H_2$, each describing an individual NH-SSH chain.

\begin{equation}
%\small
    H_1 = -i
    \begin{bmatrix}
        0 & f'+g' & & & \\[4pt]
        f'-g' & 0 & f+g & & \\[4pt]
        & f-g & 0 & f'+g' & \\[-2pt]
        & & f'-g' & 0 & \ddots\\[-1pt]
        & &  & \ddots & \ddots
    \end{bmatrix}
    \label{eq:NH-SSH Hamiltonian}
\end{equation}

Here the parameters are dressed as $g=\bar{t}+t_0 \,,\, g'=\bar{t}-t_0 \,,\, f=\bar{\gamma}+\gamma_0 $ and $f'=\bar{\gamma}-\gamma_0$. 
The Hamiltonian $H_2$ corresponding to the other chain can be easily obtained by interchanging the primed parameters $(g',f')$ with their unprimed counterparts $(g,f)$. 

\begin{figure*}
    \centering
    \includegraphics[width=\linewidth]{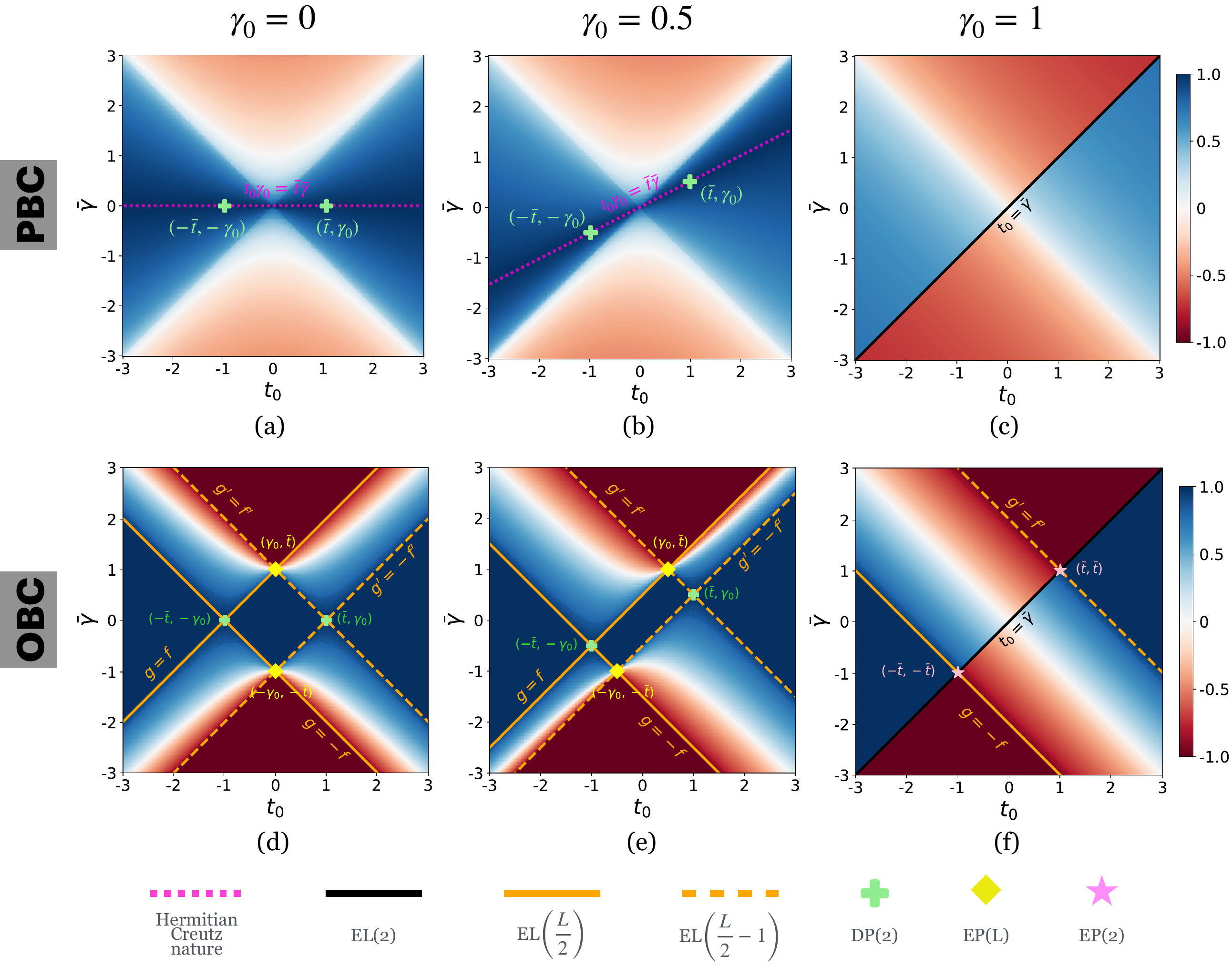}
    \caption{
            \small
            \justifying
            \textbf{Spectral phase diagrams of the non-Hermitian Creutz ladder.}
            Color maps of the spectral density measure $M$ under periodic (top row) and open (bottom row) boundary conditions in the $(t_0,\bar{\gamma})$ parameter space for three different values of $\gamma_0=(0,0.5,1)$ along the three columns for $\bar{t}=1$ and system size $L=50$. Blue (red) regions correspond to entirely real (imaginary) spectra, while color gradients indicate complex spectra. Orange lines denote exceptional lines where the Hamiltonian becomes defective and the similarity transformation fails. Green crosses mark diabolical-point flat bands, while black solid lines indicate exceptional flat bands. Yellow diamonds highlight junction points where real, imaginary, and complex spectral regions meet, forming triple-point-like structures. The figure reveals boundary condition dependent spectral transitions, including real–complex transitions mediated by diabolical points rather than exceptional points.
            }
    \label{fig:spectral domain numerical}
\end{figure*}
\textit{OBC Spectrum}. 
Unlike Hermitian systems, the bulk physics—and hence the spectrum—of a non-Hermitian chain under OBC is not captured by its PBC Hamiltonian, reflecting the breakdown of conventional BBC in non-Hermitian systems. 
For the present model, however, it is possible to perform a non-unitary similarity transformation on the OBC NH-SSH Hamiltonian that restructures it into a Hermitian SSH Hamiltonian. 
This procedure is known as an imaginary gauge transformation (IGT) \cite{hn1igt1,hn2igt2,igt3}.
The similarity matrix $S_1$ we employ for transforming $H_1$ has the form

\begin{equation}
\begin{aligned}
    &S_1 = \text{diag} \left( j_1,l_1,j_2,l_2, \cdots, j_n, l_n, \cdots, j_{\frac{L}{2}},l_{\frac{L}{2}}\right)
    \\
    &j_1 = 1, \quad l_n=\sqrt{\frac{f'+g'}{f'-g'}} \, j_n, \quad j_{n+1} = \sqrt{\frac{f+g}{f-g}}\,l_{n}
\end{aligned}
\label{eq:S-IGT}
\end{equation}

From the transformed Hamiltonian we get the OBC bulk spectrum in the thermodynamic limit to be

\begin{equation}
    E^{\text{OBC}}_\pm (q) = \pm \sqrt{u^2+v^2+2uv\cos q}
    \label{eq:OBC spectrum main}
\end{equation}

where, $q=[0,2\pi)$. Here the new parameters for inter cell hopping is $u = \sqrt{g^2-f^2}$ and intra cell hopping is $v=\sqrt{g'^2-f'^2}$. A similar IGT on the other NH-SSH chain yields a Hamiltonian with the roles of $u$ and $v$ interchanged, resulting in the same dispersion relation.

Despite the dispersion relation having the familiar form of a Hermitian SSH model, the energy values can in general be complex since the effective hopping parameters $u$ and $v$ may themselves be complex. 
When both $u$ and $v$ are real (imaginary) the spectrum is completely real (imaginary), otherwise it is complex. 

This is better visualized in \cref{fig:spectral domain numerical} where we have plotted a map of the \textit{spectral density function} $M$ \cite{sdf1}, in the parameter space $(t_0,\bar{\gamma})$.
The quantity $M$ serves as an  indicator, taking the value 1 (-1) when the entire spectrum is real (imaginary), and intermediate values when the spectrum is complex or mixed. It is defined as

\begin{equation}
    M = \frac{1}{2L} \sum_{\{E\}} \left( |\cos \theta|-|\sin \theta| \right)
    \label{eq:M,spectrum indicator 2}
\end{equation}
with $\theta = \arg (E)$ and $\{E\}$ being the energy spectrum. The plots in figure \ref{fig:spectral domain numerical}(d-f) clearly show that the entire parameter domain is divided into three parts based on its OBC spectral nature— completely real (blue), completely imaginary (red) and complex or mixed (shown in colour gradient)—emphasizing the hidden pseudo-(anti-)Hermitian nature of the OBC Hamiltonian. 
The three spectral regimes are separated by four exceptional lines (ELs), defined by the conditions $u=0$ (solid orange lines) and $v=0$ (dashed orange lines), across which the spectrum goes through a transition from real or imaginary to complex. 
Along these lines the IGT fails because the similarity matrix $S$ becomes singular and the OBC chain displays exceptional dynamics.

$u=0$ corresponds to the pair of orthogonal straight lines $g=f$ and $g=-f$, shown as solid orange lines and $v=0$ corresponds to the pair $g'=f'$ and $g'=-f'$, shown as dashed orange lines in the \cref{fig:spectral domain numerical}(d,e).
At generic points along these lines, excluding their intersection points, the Hamiltonian of the full system reduces to the direct sum, $D_2(0) \oplus J_{\frac{L}{2}-1}(\pm\lambda) \oplus J_{\frac{L}{2}}(\pm\lambda) $, where $D$ and $J$ represent diagonal and Jordan block matrices respectively, with the subscript denoting their dimensions and the arguments their associated eigenvalues.
Here $\lambda = \sqrt{g^2-f^2}$ along the dashed orange lines and $\lambda = \sqrt{g'^2-f'^2}$ along the solid orange lines. 
The block $D_2(0)$ corresponds to a pair of zero-energy edge states arising from one of the NH-SSH chains in the regime where the absolute value of inter-cell hopping exceeds the intra-cell hopping. 
Consequently, along these exceptional lines the non-Hermitian Creutz ladder becomes highly defective, supporting only four linearly independent eigenstates, all localized at the same edge of the ladder.

At the parameter values where the exceptional lines intersect, the spectral properties change qualitatively and therefore merit separate discussion. At the intersection points \((\pm \gamma_0,\pm \bar{t})\), where a solid and a dashed line meet—marked by yellow diamonds in \cref{fig:spectral domain numerical}(d,e)—the Hamiltonians \(H_1\) and \(H_2\) of the two NH-SSH chains each reduce to a single Jordan block of size \(L\) with \(\lambda=0\). As a result, the Hamiltonian takes the form \(\mathcal{H}_{\mathrm{OBC}} \cong J_L(0)\oplus J_L(0)\), and the entire OBC spectrum collapses to the origin of the complex energy plane. These points \((\pm \gamma_0,\pm \bar{t})\) thus act as \textit{triple points} in parameter space, serving as junctions where three distinct spectral regimes—real, imaginary, and complex—meet.

The remaining two intersection points \((\pm \bar{t}, \pm \gamma_0)\), marked by green crosses, correspond to diabolical points (DPs) \cite{dptp}, where the open-boundary Hamiltonian reduces to \(\mathcal{H}_{\mathrm{OBC}} \cong D_2(0)\oplus D_{L-1}(\pm 2\sqrt{\bar{t}^2-\gamma_0^2})\). At these parameter values along the exceptional lines, the Hamiltonian exhibits Hermitian degeneracies and is therefore fully diagonalizable. Consequently, any continuous path in parameter space connecting the real spectral (blue) region to the complex spectral (shaded) region through these points undergoes a real-to-complex spectral transition mediated by a diabolical point.

At $\bar{t}=\pm\gamma_0$, the OBC chain exhibits the same exceptional spectral character as the PBC system along the line $t_0=\bar{\gamma}$, shown in fig. \ref{fig:spectral domain numerical}(f), in the sense that the spectrum collapses into an EFB associated with EP2 defective eigenstates. 
Along this line the open chain Hamiltonian is composed of order two Jordan blocks, $\mathcal{H}_{\mathrm{OBC}} \cong [ \oplus J_2(0)]^L$,  except at the intersection points \((\pm\bar{t},\pm\bar{t})\) with the other ELs, where it instead takes the form $\cong D_2(0) [\oplus J_2(0)]^{(L-1)}$.
The localization properties and wave-packet dynamics of both the PBC and OBC chains at these parameter values are discussed in subsequent sections.
\label{sec:Spectrum}

%%%%%%%%%%%%%%%%%%%%%%%%%%%%%%%%%%%%%%%%%%%%%%%%%%%%%%%%%%%%%%%%%%%%
%New section
%%%%%%%%%%%%%%%%%%%%%%%%%%%%%%%%%%%%%%%%%%%%%%%%%%%%%%%%%%%%%%%%%%%%

\section{non-Hermitian skin effect}
\begin{figure*}
    \centering
    \includegraphics[width=\linewidth]{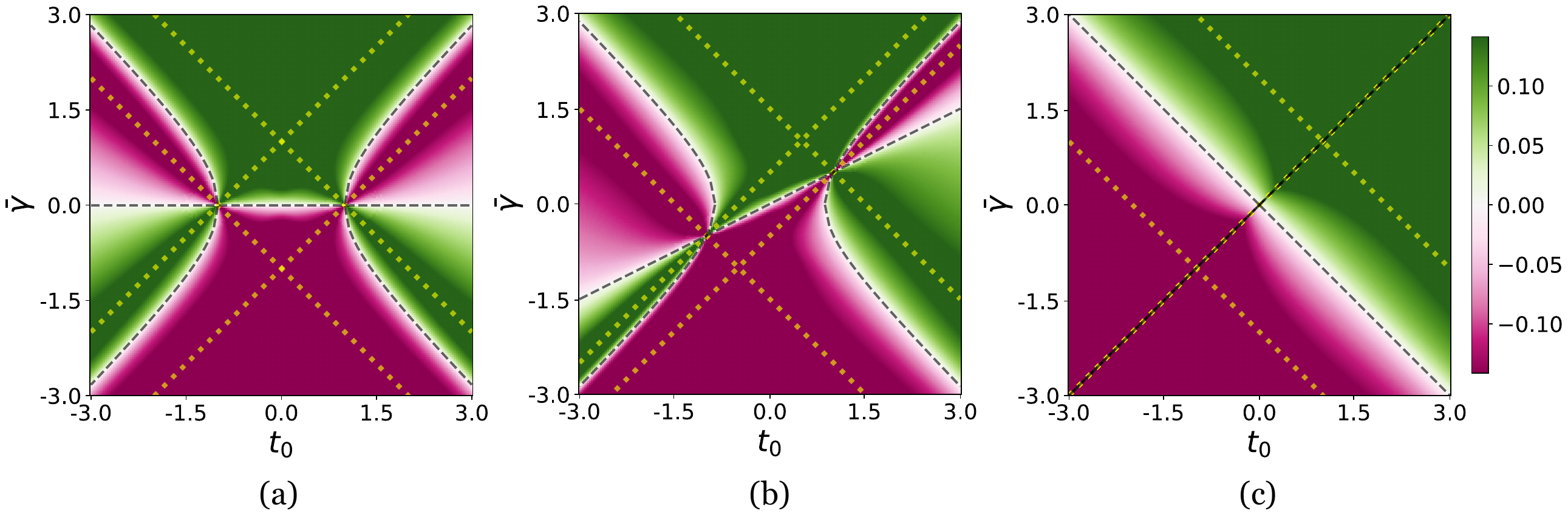}
    \caption{
        \small
        \justifying
        \textbf{NHSE localization domains.}
        Colour map of $<$dIPR$>$ in the $(t_0, \bar \gamma)$ plane. Green (pink) regions correspond to left- (right-) localized bulk eigenstates, while white regions indicate extended states where the bulk–boundary correspondence is restored. Dashed black curves denote parameter values where the similarity transformation becomes unimodular, and dotted yellow lines indicate exceptional lines. The localization domains correlate strongly with the spectral topology and exceptional structures shown in \cref{fig:spectral domain numerical}. In (a) and (b) the black and yellow lines meet exactly at the DPs where eigenstates are compactly localized. On the other hand in (c) the overlap of the two lines suggest for an entire line of such localized states.
        }
    \label{fig:dIPR}
\end{figure*}

Non-Hermitian lattice systems admit two distinct and widely discussed notions of topology. One is band topology, which concerns topological invariants associated with the band structure and leads to zero-energy edge states under open boundary conditions. The other is spectral topology, which is defined through the winding of the complex energy spectrum and is closely tied to the non-Hermitian skin effect (NHSE) \cite{NHSE2,NHSE8}, where an extensive number of bulk states localize at system boundaries.

In this work, we restrict our attention to spectral topology, which in non-Hermitian systems manifests as boundary localization of bulk eigenstates under open boundary conditions. The analytical tractability of the model allows these localization properties to be read off directly from the structure of the open-boundary eigenstates themselves. At the same time, numerical diagnostics offer an intuitive way to visualize how localization evolves across parameter space. We therefore combine explicit analytical results with numerical measures of localization to characterize both the direction and the strength of the non-Hermitian skin effect.

To identify and categorize the strength and polarity of the NHSE of the eigenstates we use the conventional \textit{directional inverse participation ratio} (dIPR), defined through

\vspace{-5pt}
\begin{equation}
    \text{LIPR}_n = \frac{\sum\limits_{j=1}^{L/2} (|\psi_{n,j}^{a}|^4 + |\psi_{n,j}^{b}|^4) }{\braket{\psi_n}{\psi_n}^2}
\end{equation}
\vspace{-5pt}
\begin{equation}
    \text{RIPR}_n = \frac{\sum\limits_{j=L/2+1}^{L} (|\psi_{n,j}^{a}|^4 + |\psi_{n,j}^{b}|^4) }{\braket{\psi_n}{\psi_n}^2}
\end{equation}
\vspace{-5pt}
\begin{equation}
    \text{dIPR}_n = \text{LIPR}_n - \text{RIPR}_n
\end{equation}
\vspace{-5pt}
\begin{equation}
    \expval{\text{dIPR}} = \frac{1}{2L} \sum_{n=1}^{2L} \text{dIPR}_n
    \label{eq:dIPR}
\end{equation}

The $\text{LIPR}_n$ ($\text{RIPR}_n$), defined as the aggregate of the fourth power of the absolute value of the n-th eigenstate over the left (right) half of the ladder. The $\expval{\text{dIPR}}$ is the difference of the left and right IPRs averaged over all eigenstates. This scalar quantity makes it easy to visualize the nature of NHSE over a parameter domain. IPRs are defined to have values inverse of the localization length ($\xi^{-1}$) of the eigenstates. So larger values correspond to stronger localization, whereas extended states exhibit values that vanish in the thermodynamic limit. Positive (negative) value will imply left (right) localization while values near zero will indicate extended states.

Figure \ref{fig:dIPR} shows the plots for the $\expval{\text{dIPR}}$ at three different values of $\gamma_0$. The pink and green regions indicate right- and left-localized eigenstates, respectively, while the intervening white regions correspond to vanishing dIPR and signal the presence of extended states. This numerical measure provides a direct visualization of the boundary localization of eigenstates and allows the parameter space to be classified according to the direction and strength of localization.

The origin of this boundary localization can be traced back to the IGT of \cref{eq:S-IGT} used to diagonalize the open-boundary NH-SSH chains. 
The structure of the similarity matrix implies that the eigenvectors will be exponentially localized at either end of the ladder. From this transformation the growth/decay factor of the eigenstates is found to be $\sqrt{(f+g)(f'+g')/(f-g)(f'-g')}$, which gives an inverse localization length of $\xi^{-1} = \frac{1}{2}[\log (f+g)+\log(f'+g')-\log(f-g)-\log(f'-g')]$.

The BBC is restored when the PBC spectrum does not enclose a finite area in the complex energy plane and coincides with the OBC spectrum. 
For pseudo-(anti) Hermitian systems where $S$ matrix maps the original Hamiltonian to a (anti) Hermitian one, the recovery of BBC depends on the boundedness of this transformation. 
In our system this correspondence to the growth factor of the eigen states being unimodular, leading to the conditions $t_0^2+\gamma_0^2=\bar t^2 +\bar\gamma^2$ and $t_0\gamma_0=\bar t \bar \gamma$, corresponding to a hyperbola and a straight line in the parameter space of $t_0,\bar{\gamma}$, shown with dashed black lines in \cref{fig:dIPR}. The spectral nature of the system for two generic points along these curves are shown in \cref{fig:spectrum plots}(e,f). 
Along these curves the eigenvectors of the system are extended over the bulk of the chain, except at the points where the hyperbola and the straight line intersect which are the DPs of the parameter space with compactly localized eigenstates.
\label{sec:NHSE}

%%%%%%%%%%%%%%%%%%%%%%%%%%%%%%%%%%%%%%%%%%%%%%%%%%%%%%%%%%%%%%%%%%%%
%New section
%%%%%%%%%%%%%%%%%%%%%%%%%%%%%%%%%%%%%%%%%%%%%%%%%%%%%%%%%%%%%%%%%%%%

\section{Wave packet dynamics}
\begin{figure*}
    \centering
    \includegraphics[width=\linewidth]{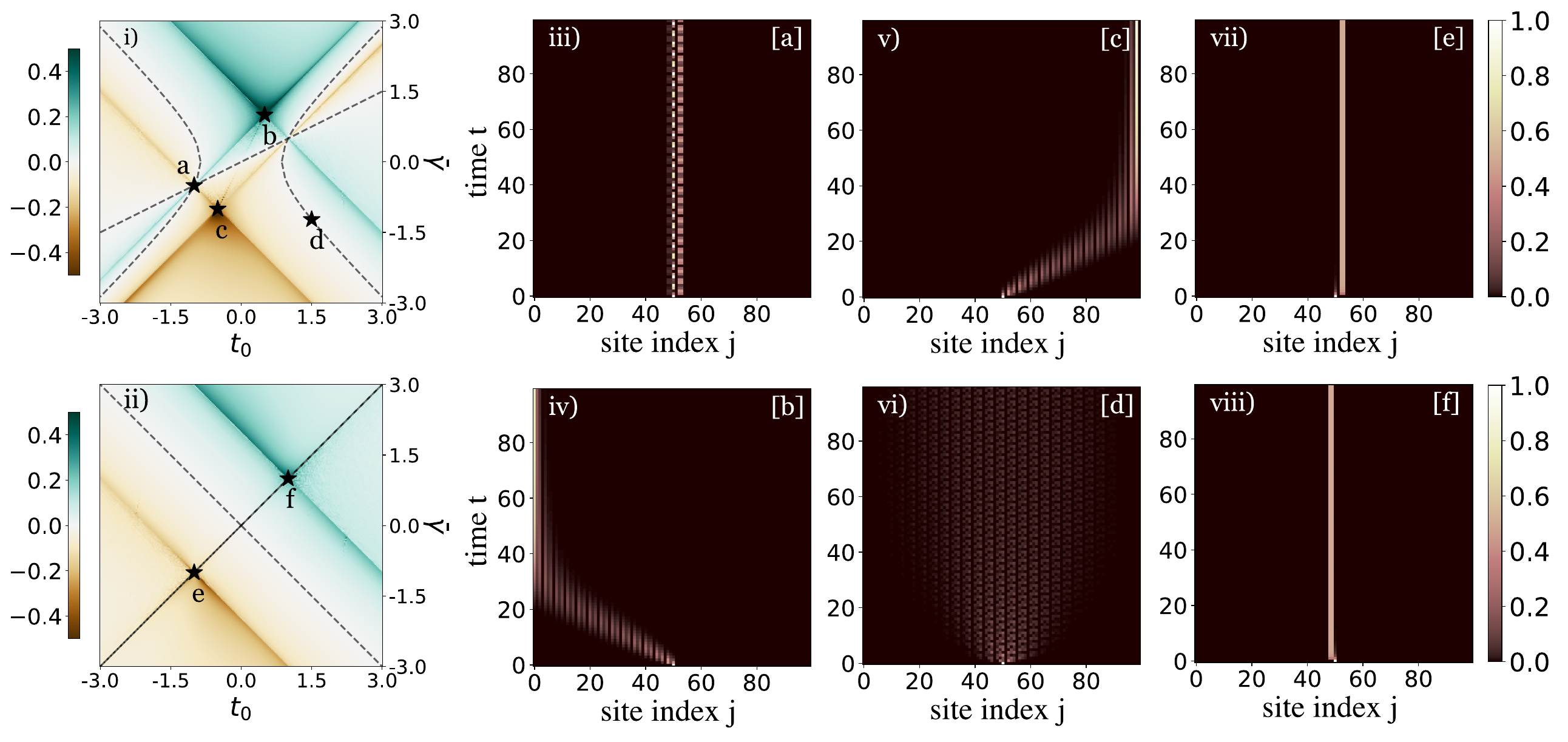}
    \caption{
    \small
    \justifying
    \textbf{Wave packet evolution and compacton dynamics.}
    (i,ii) Color maps of the modified IPR (mIPR) characterizing the long-time dynamics of a wave packet initially localized at the center of the ladder for two different values of $\gamma_0=(0.5,1)$ on a system of size $L=50$. Green (brown) regions indicate left (right) dynamical localization. Panels (iii–viii) show time evolution of wave-packet intensity for representative parameter values marked in panels (i,ii). At diabolical-point flat bands, the wave packet exhibits oscillatory compacton dynamics, while at exceptional flat bands the dynamics becomes non-oscillatory and more stringently localized. Along exceptional lines, dynamical localization mirrors the static non-Hermitian skin effect, whereas at BBC restoration points the wave packet spreads symmetrically.
    }
    \label{fig:mIPR-wave dynamics}
\end{figure*}

We now examine the time dynamics of wave packets in the non-Hermitian Creutz ladder. An initially localized wave packet is placed at the center of the ladder and evolved in time under open boundary conditions. Owing to the non-Hermitian skin effect, we expect the evolving wave packet to exhibit directional localization determined by the polarity of the underlying eigenstates. To quantify this behavior, we introduce a modified directional inverse participation ratio (mIPR),
\begin{equation}
\text{mIPR} = \sum_{j=1}^{L} \frac{(L/2-j)}{L/2}
\left(|\psi_{j,a}|^4 + |\psi_{j,b}|^4\right),
\label{eqn:mIPR}
\end{equation}
which measures the left--right asymmetry of the time-evolved state relative to the center of the lattice. A positive (negative) mIPR indicates localization toward the left (right) boundary, while values near zero correspond to unbiased spreading or central localization.

Figures \ref{fig:mIPR-wave dynamics}(i,ii) show the mIPR map for the open chain at two representative values of \(\gamma_0=0.5,1\). The close correspondence between these plots and the NHSE phase diagrams in \cref{fig:dIPR}(b,c) demonstrates that the directionality of the evolving wave packet is governed by the polarity of the right eigenstates. In regions where bulk--boundary correspondence is restored, the wave packet spreads symmetrically about its initial position, resembling Hermitian dynamics [\cref{fig:mIPR-wave dynamics}(vi)]. By contrast, along exceptional lines where the eigenstates accumulate at one boundary, the wave packet evolves toward the same edge [figs \ref{fig:mIPR-wave dynamics}(iv,v)].

A defining feature of the Hermitian Creutz ladder is compacton dynamics at flat-band points, where wave packets remain confined to a small number of plaquettes due to destructive interference. We find that such compact localization persists in the non-Hermitian Creutz ladder. At diabolical flat-band points, the wave packet exhibits oscillatory compacton dynamics, periodically spreading between neighboring cells [\cref{fig:mIPR-wave dynamics}(iii)], albeit with asymmetric weight distribution induced by non-Hermiticity.

Along the exceptional flat-band line defined by \(\bar{t}=\gamma_0\) and \(t_0=\bar{\gamma}\), the compacton dynamics is qualitatively modified. Here the wave packet remains spatially confined but no longer exhibits oscillatory revival. At the intersection points of the exceptional lines with the exceptional flat-band line, the dynamics becomes strongly skewed: at long times the wave packet occupies only a single lattice cell adjacent to its initial position [figs. \ref{fig:mIPR-wave dynamics}(vii,viii)]. Since these compacton dynamics originate from local second-order degeneracies---either Hermitian (\(D_2\)) or exceptional (\(J_2\))---they are insensitive to boundary conditions. Consequently, the same dynamical behavior is observed under periodic boundary conditions.
\label{sec:wave dynamics}
%
%%%%%%%%%%%%%%%%%%%%%%%%%%%%%%%%%%%%%%%%%%%%%%%%%%%%%%%%%%%%%%%%%%%%
%New section
%%%%%%%%%%%%%%%%%%%%%%%%%%%%%%%%%%%%%%%%%%%%%%%%%%%%%%%%%%%%%%%%%%%%
\vspace{-10pt}
\section{Conclusion}
In this work, we presented an analytical study of a non-Hermitian extension of the Creutz ladder with generic non-reciprocal hopping, focusing on the fate of flat-band physics under non-Hermiticity and the role of boundary conditions. By exactly diagonalizing the model under both periodic and open boundary conditions, we demonstrated that flat bands persist in the non-Hermitian regime in multiple forms, including Hermitian diabolical-point flat bands and exceptional flat bands, each associated with distinct spectral and dynamical characteristics. The mapping of the ladder onto two decoupled non-Hermitian SSH chains enabled a transparent analytical understanding of these features and provided direct access to the boundary-sensitive structure of the spectrum.

A key outcome of our analysis is the explicit organization of the parameter space into regions with qualitatively different spectral character under open boundary conditions, including real, imaginary, and complex spectra, separated by exceptional lines. The strong sensitivity of the spectrum to boundary conditions highlights the breakdown of conventional bulk–boundary correspondence in this system and clarifies how non-Hermitian effects reorganize flat-band degeneracies in a ladder geometry constrained by interference. Beyond spectral properties, we showed that these distinctions manifest dynamically, with wave-packet evolution and compact localization exhibiting markedly different behavior depending on whether flat bands arise from diabolical or exceptional degeneracies.

Our results establish the non-Hermitian Creutz ladder as an analytically tractable platform for exploring the interplay between flat-band physics, exceptional degeneracies, and boundary-condition-dependent phenomena. The framework developed here can be extended to more general ladder geometries, imbalanced hopping configurations, or systems with interactions and disorder, and may serve as a useful guide for experimental realizations of non-Hermitian flat-band systems in photonic, mechanical, or circuit-based platforms. 

\label{sec:conclusion}

\section*{acknowledgments} 
SC thanks DST, India for support through SERB project SRG/2023/002730 and DST, India, for support through the project DST/FFT/NQM/QSM/2024/3.

%%%%%%%%%%%%%%%%%%%%%%%%%%%%%%%%%%%%%%%%%%%%%%%%%%%%%%%%%%%%%%%%%%%%
%New section
%%%%%%%%%%%%%%%%%%%%%%%%%%%%%%%%%%%%%%%%%%%%%%%%%%%%%%%%%%%%%%%%%%%%

\bibliographystyle{apsrev4-2}
\bibliography{Bibliography}

\end{document}